\documentclass{PoS}
\bibliography{PoS.bst}
\newcommand{\grs}{GRS~1915+105~}
\newcommand{\cygc}{Cyg~X-3~}

\newcommand{\sgr}{V4641~Sgr~}
\title{Relativistic superluminal radio jets in microquasars in our galaxy }

\ShortTitle{Relativistic superluminal radio jets in microquasars}

\author{\speaker{J S Yadav}%
         \\
        TATA Institute of Fundamental Research, Mumbai, India\\
        E-mail: \email{jsyadav@crspa.tifr.res.in}}

\abstract{  
    We discuss the origin of  superluminal radio jets in Black hole X-ray 
binaries with relativistic radio jets in our Galaxy popularly known as 
microquasars. We classify the relativistic superluminal jet  according  to 
the radio emission in black hole X-ray binaries (transient or persistent) 
rather than the mass of the companion. The black hole X-ray binaries with
transient radio emission (mostly LMXBs) produce  superluminal jets
with $\beta_{app} >$ 1  when the accretion rate, $\dot{m}_{accr}$ is high
and the bolometric luminosity L$_{bol}$  approaches the Eddington
Luminosity, L$_{Edd}$. On the other hand, the black hole X-ray binaries
with persistent radio emission (mostly HMXBs) produce superluminal jets
with $\beta_{app} <$ 1 at
relatively low  $\dot{m}_{accr}$. We
specially discuss the case of \sgr, a HMXB with transient radio emission
which produces superluminal radio jets like  in LMXBs.}
\FullConference{VI Microquasar Workshop: Microquasars and Beyond\\
		 September 18-22, 2006\\
		 Como, Italy}

\begin{document}
\section{Introduction}
The Black hole X-ray binaries with relativistic  radio jets have earned the
name ``microquasars'' as they are stellar-mass analogs of  the massive
extragalactic black hole systems in Quasars and Active Galactic Nuclei
(AGNs) \cite{mira94,ting95}.  The total number of X-ray binaries
(HMXBs plus LMXBs) as per  the  recent catalogues is 280 which
includes neutron star and black hole X-ray binaries
\cite{liu00,liu01}. Most of the LMXBs
are transients while most of the HMXBs are persistent.
The current number of microquasars  is $\sim$ 14 which
are listed in Table~\ref{microq} out of total 43 X-ray binaries in
which radio emission is detected \cite{mccl04, par05}.

\begin{table}[b]
\begin{center}
\scriptsize
\begin{tabular}{|lccccc|}
\hline
\hline
Source &Type of &D &M$_{compact}$&Radio &$\beta_{appr}$\\
 &binary & (kpc) &M$_{\odot}$ &emission&\\
\hline
LS~I~+61~303&HMXB&2.0&--&p&$\ge0.4$\\
V4641~ Sgr&HMXB&$\sim$10&9.6&t&$\ge$9.5\\
LS~5039&HMXB&2.9&3(?)&p&$\ge$0.15\\
SS~433&HMXB&4.8&11$\pm$5(?)&p&0.26\\
Cygnus X-1&HMXB&2.5&10.1&p&--\\
Cygnus X-3&HMXB&9&--(?)&p&0.69\\
XTE~1118+480&LMXB&1.9&6.9$\pm$0.9&t&--\\
XTE~J1550-564&LMXB&5.3&9.4&t&$>$2\\
GRO~J1655-40&LMXB&3.2&7.02&t&1.1\\
GX~339-4&LMXB&$>$6&5.8$\pm$0.5&t&$\ge$2.2\\
1E~1740.7-2942&LMXB&8.5(?)&--&p&--\\
XTE~J1748-288&LMXB&$\ge$8&$>$4.5(?)&t&1.3\\
GRS~1758-258&LMXB&8.5(?)&--&p&--\\
GRS~1915+105&LMXB&12.5&14$\pm$4&t&1.2--1.7\\
\hline
\end{tabular}
\caption{Microquasars in our Galaxy. Radio jet sources with neutron 
stars like Circinus~ X-1 and Scorpius~X-1 are not included.}
\label{microq}
\end{center}
\end{table}

Mirabel \& Rodriguez (1994) discovered first microquasar \grs in our galaxy  
with superluminal jets. This microquasar shows exceptionally
high variability in both X-rays and radio 
\cite{muno99,yada99,bell00}.
All the radio flares observed in \grs  can be
  broadly put into two groups on the basis of their flux, radio 
  spectrum and spatial distribution; (1) the superluminal  flares
  (200--1000 mJy) which have steep radio spectra and are seen at 
  large distances ($\ge$ 240 AU), and (2) all other flares (5--360 mJy) 
  which include the preplateau flares, radio oscillations \& 
  discrete flares  and the steady radio flares 
  during the plateaux. All these flares  have flat or inverted 
  radio spectra
  and are observed close to the compact object ($<$ 200AU).
  These radio flares can be understood as mass ejections of adiabatically 
  expanding  self absorbing synchrotron  clouds 
  \cite{mira98,eike98,yada01,ishw02}.
  
  The physical connection between X-ray emission  and the superluminal flares
  has been  the hardest to understand\cite{fend04b,fend04}.
 Recently, Yadav (2006) has studied X-ray properties during plateaux  
 and the following 
 superluminal radio jets in \grs and has 
 provided tight correlation between the accretion disk and
  the radio jet parameters. 
  It is suggested that the superluminal jets are due to
  the internal shock which  forms when the oscillation or discrete baby jet 
  (compact jet) catches up and interact with the previously generated slowly 
  moving wind from the accretion disk (see next section for details).
     We attempt in this paper to broaden 
  this understanding of the 
  superluminal radio jets
  to  HMXBs with mostly persistent radio emission in which the wind from the
  companion is the usual source of accretion.  \cygc is the most active 
  in this class of HMXBs. This source was first detected 
  in 1967 \cite{giac67},  
  and is known to produce  huge superluminal radio
  jets \cite{mill04,miod01}.
We have studied X-ray properties during the plateaux \& the class
$\beta$ in \grs and during the low hard state in \cygc preceding 
superluminal flares. We also studied radio properties of superluminal
and discrete/oscillation jets (see Table~\ref{radiox}).
 For X-ray spectral analysis, we have used REXTE PCA/HEXTE
 data in the range 4--150 keV.  
 The decay time constant of a superluminal flare is 
calculated by fitting
exponential decay profile to the  2.25 GHz GBI 
radio monitoring data. Yadav (2006)  has given  details of 
X-ray and radio data analysis.
\begin{table}[t]
\begin{center}
\scriptsize
\begin{tabular}{|llll|llll|}
\hline
\hline
\multicolumn{4}{|c|}{Radio Flare Properties }  &\multicolumn{4}{|c|}{Associated X-ray properties} \\
\hline 
MJD &Source&Type of&Flux(MJy)&MJD/Day&X-ray&N$_H$ &$\chi^2_{\nu}$(dof)\\
 & & flare  &2.25 GHz &&Flux$^a$ &(10$^{22}$ cm$^{-2}$) &(dof)  \\
\hline
50915 & \grs & Superluminal& 920 &50913/1998 Apr. 11 &2.14 &14.95$^{+0.82}_{-0.65}$  &0.83(94) \\
52105 & \grs & Superluminal& 368 & 52101/2001 Jul. 11 & 1.61 &10.60$^{+0.77}_{-0.40}$& 0.98(87) \\
50674& \grs &Oscil./baby jets$^b$ &60-80$^c$ &50674/1997 Aug. 14 & 1.00& 7.0$^d$ &0.92(75) \\
50700  &\grs &Oscil./baby jets  &60-80$^c$ &50700/ 1997 Sep. 9 &1.01  &7.0$^d$ & 0.71(75) \\
50610  &\cygc &Superluminal &4000 &50605/ 1997 Jun. 5 &0.76  &28.91$^{+1.50}_{-0.40}$ &0.99 (70)  \\
51654& \cygc & Superluminal& 13000& 51646/ 2000 Apr. 12& 0.45$^e$&25.53$^{+14.00}_{-7.00}$& 0.83(64) \\
\hline
\multicolumn{8}{l}{$^a$ Integrated 4--150 keV X-ray flux in 10$^{-8}$ ergs cm$^{-2}$ s$^{-1}$} \\
\multicolumn{8}{l}{$^b$ Radio flux is calculated from IR flux after dereddening,
$^c$ Radio flux at 8.3 GHz} \\
\multicolumn{8}{l}{$^d$ This value is for fixed N$_H$. This is also consistent with variable N$_H$ which results 8.40$^{+1.50}_{-1.45}$} \\
\multicolumn{8}{l}{$^e$ Only 20 min RXTE PPC/HEXTE data and both the total 
X-ray flux and N$_H$ are likely to be underestimated}\\
\end{tabular}
\caption{All selected superluminal radio  jets, oscillation/baby jets  and their X-ray Properties from RXTE PCA/HEXTE data}
\label{radiox}
\end{center}
\end{table}
%
\section{Results and discussion}
	We list  in Table~\ref{microq} data of all the known microquasars.
We have limited our list to only black hole X-ray binaries.
Most of the HMXBs in Table~\ref{microq} show
persistent radio emission  while  most of the LMXBs  
 are transient in nature. It is
clear from Table~\ref{microq} that all the microquasars with transient 
radio emission produce superluminal jets with $\beta_{appr} >$ 1 while
those with persistent radio emission produce superluminal jets with 
$\beta_{appr} <$ 1. 
\begin{figure}
\begin{center}
\includegraphics[angle=270,width=.8\textwidth]{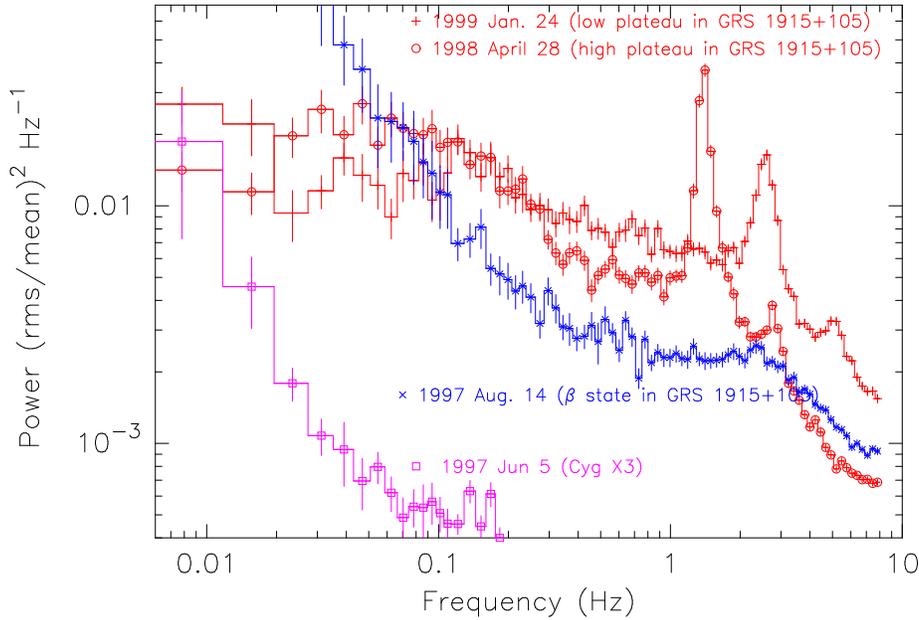}
\caption{Normalised power density spectra in 0.007 -- 9 Hz frequency      
range observed during the plateaux state (1998 April 28 and 1999 January 24)
and  during the X-ray class $\beta$ on 1997 August 14 in \grs. The PDS
observed during the low hard state on 1997 June 5 in \cygc preceding 
superluminal flare.  }
\label{pdsboth}
\end{center}
\end{figure}
  
Yadav (2006)  has  shown  that  the  accretion rate, 
$\dot{m}_{accr}$ is very high during the radio plateaux in \grs and 
the bolometric luminosity L$_{bol}$ approaches Eddington Luminosity L$_{Edd}$.
It is suggested that such luminous accretion disk during the plateaux 
always accompany with radiation-driven wind.   The internal 
  shock  forms in the previously generated slowly moving
  wind from the accretion disk during the plateaux with $\beta$ $\le$ 0.01  
  as the fast moving 
  discrete jet with $\beta$ $\sim$ 1  catches up and interacts with it
  \cite{yada06, kais00}.  Both the components; 
  slow moving wind and fast moving jet are related to the accretion disk 
  during plateau state and the strength \& speed of these two  should 
  determine the power of the internal shock and of the superluminal radio
  jet. 
A tight correlation  is observed between  peak 
flux of superluminal flares  and  absorption column density, N$_H$ 
with correlation coefficient of 0.99 (see our other paper in this proceeding)
which suggests that for wind strength corresponding to  
N$_H \le$ 8.3$\pm$1.5$\times$10$^{22}$ cm$^{-2}$, no superluminal jet
will  be produced.
The superluminal radio jets in most of LMXBs are produced when 
the source is in 
very high luminous state (VHS) like plateaux in \grs a prerequisite
to produce slow moving strong  wind.
The evidences for a Seyfert-like warm absorber have been found in 
many LMXBs like  GRO J1655-40, GX 339-4 and XTE J1650-500 beside \grs 
\cite{mill04b,ueda98,kota00}.
The
$\beta_{appr}$ of observed superluminal jets in these systems 
is $>$ 1 and peak radio
flux is in the range of 0.4 -- 2 Jy. The rise time of superluminal 
flares  in these sources is in the range 0.25 -- 0.5 day \cite{yada06,
fend04}.
   
The absence of superluminal jets during
  the class $\beta$ in \grs is attributed to the absence of wind \cite{yada06}.
In Table~\ref{radiox}, we give details of radio flares and the associated
X-ray properties during the plateaux and the class $\beta$ in \grs.   
  During the X-ray class $\beta$ in \grs, the total X-ray flux
  is almost half than that observed during the plateaux and the calculated
  N$_H$ is below the critical N$_H$ discussed above to produce superluminal
  jets. 
 We also present in Table~\ref{radiox} the radio properties  of two  
  superluminal flares observed in \cygc   along with associated X-ray
  properties during the preceding low hard state. The N$_H$ is  in the
  range of 25 --30$\times$10$^{22}$ cm$^{-2}$ which is consistent with
  the  strong wind from the companion. The total X-ray flux  is
  close to that observed during the class $\beta$ in \grs. The observed 
  N$_H$ and the 
  X-ray flux in \cygc will suggest that the   discrete/oscillation  compact
  jet may  interact with  the wind which originates from the companion
  to produces superluminal jets in HMXBs with persistent radio emission.
  
  Figure~\ref{pdsboth} shows PDS spectra observed during low plateau 
  on  1999 January 24 and during high plateau on 1998 April 28 in \grs. 
  Yadav (2006) has  suggested that the fast
  variability is suppressed by photon scattering in the enhanced  wind on
  1998 April 28 and hence reducing the power in the PDS at higher frequencies
  (see Paper 1 for detailed discussion).
We also show in Figure 1 the calculated PDS for class $\beta$ in \grs and 
during low hard state in  \cygc.
The shape of PDS during class $\beta$ in \grs is similar to the shape of PDS
observed in \cygc except the PDS power in \cygc  at higher frequency is highly 
suppressed due to dense 
wind from the companion \cite{chou04}. 
The depleted IR emission in \cygc as compared to class $\beta$ in 
\grs also indicates the
presence of strong wind in \cygc \cite{yada06}.

\begin{figure}
\begin{center}
\includegraphics[angle=270,width=.6\textwidth]{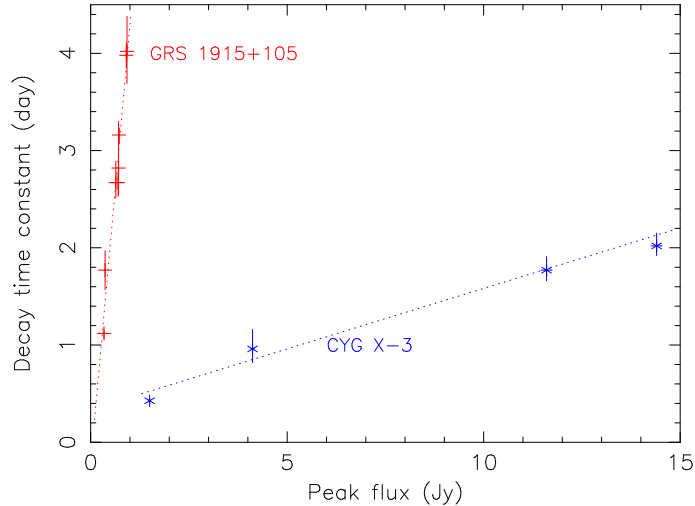}
\caption{The peak flux of the superluminal flares as a function of 
decay time constant of the superluminal flares observed in \grs and \cygc
from the analysis of 2.25 GHz GBI data.}
\label{peakcon}
\end{center}
\end{figure}

In the case of HMXBs with persistent radio
emission  like \cygc, superluminal jets are produced  at relatively low 
$\dot{m}_{accr}$ as the slow moving medium; wind is provided by 
the companion. In this case
power of superluminal jets is coming from the accretion disk as well as  
from the wind which originates from the companion.  
The superluminal flare (transient)  mode
boosts the radio luminosity over the X-ray luminosity 
in microquasars and the boosting of
radio luminosity is higher in \cygc than that in \grs \cite{nipo05}. This
is because the wind from the companion provide additional power 
to superluminal jets in \cygc. 
 Figure~\ref{peakcon} shows decay time constant
as a function of peak flux of superluminal flares observed in \grs and
\cygc obtained from the analysis of 2.25 GHz GBI monitoring data. The 
slow wind and the compact jet both originate from the accretion disk in \grs 
while the wind from the companion provides accreting material as well as slow 
moving medium  to evolve internal shock in \cygc and hence both 
follow linear rise. The \grs data represent superluminal jets produced 
 when $\dot{m}_{accr}$ is very high and the L$_{bol}$ approaches 
L$_{Edd}$ while \cygc data represent superluminal jets produced at 
relatively low $\dot{m}_{accr}$  but with dense wind from the companion.
Superluminal jets from other microquasars should fall between these two. 

The case of \sgr is special as it is a HMXB with transient
radio emission  and has produced superluminal jets with 
$\beta_{appr} \ge$ 9.5
(Table~\ref{microq}) \cite{hjel00}.  The giant outburst in
1999 September 14-15 is attributed  to an episode of super-Eddington
accretion  onto the black hole  and the L$_{bol}$ approaches
L$_{Edd}$
\cite{revn02}. During this outburst, extended
 optically thick envelope/outflow has been seen. Optical outbursts 
 have been seen
 earlier in GRO~J1655-40 and XTE~J1550-564, both of them LMXBs with transient
 radio emission. During the outburst, wind is reported and the N$_H$  has
 increased substantially which is similar to what we observe during the
 plateaux in \grs \cite{revn02,yada06}. However these plateau like 
 conditions lasted for short duration less than a day (1999 September 
 14.9--15.7). The elongated radio jet with steep radio spectra was seen 
 on September 16 with peak flux $\sim$ 400 mJy which disappeared quickly
 \cite{hjel00}.
 This would suggest a partially developed internal shock due to lack of
 slow moving medium; the wind. It suggests that \sgr behaves like 
 LMXBs with transient radio emission.

\acknowledgments
The author  thanks  RXTE PCA/HEXTE and NSF-NRAO-NASA Green Bank Interferometer 
teams for making 
their data publicly available. 

\end{document}